%Paper: hep-th/9407084
%From: kibler@lyolav.in2p3.fr
%Date: Fri, 15 Jul 94 13:20:28 +0200

\magnification\magstep1
\baselineskip = 0.5 true cm
\parskip=0.1 true cm

  \def\sa{\vskip 0.30 true cm}
  \def\sb{\vskip 0.60 true cm}
  \def\sc{\vskip 0.15 true cm}

  \pageno = 0

  \vsize = 22 true cm
  \hsize = 15 true cm

\font\msim=msym10
\def\gr{\hbox{\msim R}}

\def\grn{\hbox{\msim N}}
\def\grz{\hbox{\msim Z}}

%   \font\logo=logoipnl scaled \magstep4
   \line{\vbox{\hsize 3.5 true cm
   \noindent
%   \logo
 P}\hfill \vbox{\hsize 3 true cm
   \noindent
   \null\hfill \bf LYCEN/9358\break
   \null\hfill November 93}}

\sc
\sb
\sc
\sb

\centerline {\bf MISCELLANEOUS PHYSICAL APPLICATIONS}

\centerline {\bf OF QUANTUM ALGEBRAS$^*$}

\sa
\sb
\vskip 0.5 true cm

\centerline {Maurice Kibler}

\sa

\centerline {Institut de Physique Nucl\'eaire de Lyon}
\centerline {IN2P3-CNRS et Universit\'e Claude Bernard}
\centerline {43 Boulevard du 11 Novembre 1918}
\centerline {F-69622 Villeurbanne Cedex, France}

\sa
\sa
\sa
\sb
\sb

\baselineskip = 0.7 true cm

\sa

\centerline {\bf Abstract}

\sb

\noindent Some ideas about phenomenological applications of quantum
algebras to physics are reviewed. We examine in particular
some applications of the algebras
$U_ q  ({\rm su}_2)$ and
$U_{qp}({\rm u }_2)$
to various dynamical systems and to atomic and nuclear
spectroscopy. The lack of a true (unique) $q$- or $qp$-quantization
process is emphasized.

\sa
\sa
\sa
\sb
\sa
\sa
\sa
\sb
\sa
\sa
\sa
\sb

\baselineskip = 0.5 true cm
\noindent $^*$ From a lecture given at the International Symposium
``Generalized Symmetries in Physics'', Arnold Sommerfeld
Institute, Technical University Clausthal, Germany
(27-29 July 1993). To appear in the proceedings of the
symposium, eds.~H.-D. Doebner, V. Dobrev and A. Ushveridze
(World Scientific, Singapore).

\vfill\eject

\vglue 0.8 true cm

\centerline {\bf MISCELLANEOUS PHYSICAL APPLICATIONS}

\centerline {\bf OF QUANTUM ALGEBRAS}

\sa
\sb
\vskip 0.5 true cm

\centerline {Maurice Kibler}

\sa
\baselineskip = 0.50 true cm

\centerline {Institut de Physique Nucl\'eaire de Lyon}
\centerline {IN2P3-CNRS et Universit\'e Claude Bernard}
\centerline {43 Boulevard du 11 Novembre 1918}
\centerline {F-69622 Villeurbanne Cedex, France}

\sa
\sa
\sa
\sb
\sb
\sa

\centerline {\bf Abstract}

\sb

\baselineskip = 0.48 true cm
\leftskip = 1.0 true cm
\rightskip = 1.0 true cm

Some ideas about phenomenological applications of quantum
algebras to physics are reviewed. We examine in particular
some applications of the algebras
$U_ q  ({\rm su}_2)$ and
$U_{qp}({\rm u }_2)$
to various dynamical systems and to atomic and nuclear
spectroscopy. The lack of a true (unique) $q$- or $qp$-quantization
process is emphasized.

\sa
\sb
\leftskip = 0 true cm
\rightskip = 0 true cm
\baselineskip = 0.59 true cm

\centerline {\bf 1. Introduction}

\sb

The concept of quantized universal enveloping algebras (or
quantum algebras) introduced in the eighties continues to
be the object of numerous developments in mathematics and
physics.
Quantum algebras may be realized in terms of $q$-deformed
bosons. The various physical applications of $q$-bosons and
quantum algebras may be naively classified in four types.

1. In a problem involving
   {\it ordinary} bosons or
   {\it ordinary} harmonic oscillators or
   {\it ordinary} angular momenta
   (orbital, spin, isospin, $\cdots$ angular momenta) or, more generally, any
   {\it ordinary} dynamical system,
one may think
of replacing them by their $q$-analogues. If the limiting case
where $q = 1$ describes the problem in a reasonable way,
one may expect that the case where $q$ is close
to $1$ can describe some fine structure effects. In this approach,
the (dimensionless) parameter
$q$ is a further fitting parameter describing
additional degrees of freedom~; the question in
this approach is to find a physical interpretation of the
(fine structure or anisotropy or curvature) parameter
$q$. Along this first type, we have the following
applications.

\item{(i)} Use of $q$-deformed oscillators for describing the
interaction between matter and radiation.

\item{(ii)} Use of $q$-deformed oscillators and application of
the quantum algebra
$U_q({\rm su}_{1,1})$ to vibrational spectroscopy of molecules.

\item{(iii)} Application of the quantum algebra
$U_q({\rm su}_{2  })$, and even
$U_q({\rm  u}_{2  })$,
to vibrational-rotational spectroscopy of molecules and nuclei.

Note that,  in connection with these utilizations and applications, we
may ask whether $q$-bosons should obey some $q$-deformed Bose-Einstein
statistics.

2. A second type of applications concerns the more general
situation where a physical problem is well described
by a given (simple) Lie algebra $g$.
One may then consider to
associate a quantized universal enveloping algebra $U_{q}(g)$
to the Lie algebra $g$.
Symmetries
described by the Lie algebra $g$ are thus replaced
by symmetries inherent to the quantum algebra $U_q(g)$.
For generic $q$ (excluding the case where $q$ is a root
of unity), the representation theory of $U_{q}(g)$
is connected to the one of $g$ in a trivial manner since
 any irreducible representation
of $g$ provides us with an irreducible representation of
$U_q(g)$.
 Here again, the
case where $q$ is close to $1$ may serve to
describe fine structure effects.

3. A third type arises by allowing the deformation parameter
$q$ not to be restricted to (real or complex) values close to
1. Completely unexpected results may result from this approach.
This is the case for instance when $q$ is a root of
unity for which case the representation theory of the quantum
algebra $U_{q}(g)$ may be very different from the one of $g$.
This may be also the case when $q$ takes values (in $\gr$ or in
$S^1$) far from unity.

4. Finally, a fourth type concerns more
fundamental applications
(more fundamental in the sense  that
the deformation parameter $q$ is not
subjected to fitting procedures).
We may mention, among
others, applications to statistical mechanics, gauge theories,
conformal field theories and so on. Also, quantum
algebras might be interesting for a true definition of the
quantum space-time.

We shall deal here with physical
applications (mainly of a phenomenological
nature) of type 1 to 3. We shall give a survey of ideas
around some protypical applications of quantum algebras
corresponding to deformations of the Lie algebra $g=A_1$.
Most of the applications have been concerned up to
now with only one parameter (say $q$). The introduction of a
second parameter (say $p$) should permit more flexibility.
Therefore, we shall briefly discuss in section 2 a particular
version of a two-parameter quantum algebra
$U_{qp}({\rm u}_2)$. Then,
we shall consider applications to~: (i) some nonrelativistic
dynamical systems, with an emphasis on the Coulomb
system which plays a so important role in atomic spectroscopy
and quantum chemistry (section 3), (ii) the classification of
chemical elements (section 4), and (iii) rotational spectroscopy
of molecules and atomic nuclei (section 5). We shall limit ourselves
in sections 2-5 to some results with a minimal
bibliography. (Further references can be obtained
from the quoted literature.) Some conclusions
shall be given in section 6.

The author thanks the organizers of the  symposium  on
``Generalized Symmetries in Physics'' for inviting him
to give this lecture.

\sa
\sb

\centerline {\bf 2. The quantum algebra $U_{qp}({\rm u}_2)$}

\sb

The quantum algebra $U_{qp}({\rm u}_2)$ can be easily
introduced in the oscillator representation [1]. Let us start by
defining the linear operators $a_+$,
                              $a_+^+$,
                              $a_-$, and
                              $a_-^+$
by the relations
$$
  \eqalign{
  a_+   \; |n_1 \rangle \otimes
           |n_2 \rangle
\; = \; &{\sqrt {[[n_1 + {1 \over 2} - {1 \over 2}]]_{qp}}}\;
|n_1 - 1 \rangle \otimes
|n_2     \rangle \cr
  a_+^+ \; |n_1 \rangle \otimes
           |n_2 \rangle
\; = \; &{\sqrt {[[n_1 + {1 \over 2} + {1 \over 2}]]_{qp}}}\;
|n_1 + 1 \rangle \otimes
|n_2     \rangle \cr
  a_-   \; |n_1 \rangle \otimes
           |n_2 \rangle
\; = \; &{\sqrt {[[n_2 + {1 \over 2} - {1 \over 2}]]_{qp}}}\;
|n_1     \rangle \otimes
|n_2 - 1 \rangle \cr
  a_-^+ \; |n_1 \rangle \otimes
           |n_2 \rangle
\; = \; &{\sqrt {[[n_2 + {1 \over 2} + {1 \over 2}]]_{qp}}}\;
|n_1     \rangle \otimes
|n_2 + 1 \rangle \cr
  }
\eqno (1)
$$
(with $a_+ | 0 \rangle \otimes
           | 0 \rangle =
       a_- | 0 \rangle \otimes
           | 0 \rangle = 0$),
where $|n_1n_2 \rangle$ is an (undeformed) vector defined on
a two-particle Fock space ${\cal F}_1 \otimes {\cal F}_2$.
In this paper, we use the notations
$$
[[X]]_{qp} = { q^X - p^X    \over q-p      } \qquad
 [X] _ q \equiv [[X]]_{qq^{-1}}
           = { q^X - q^{-X} \over q-q^{-1} }
\eqno (2)
$$
where $X$ may stand for an operator or a number.
The sets $\left\{ a_+, a_+^+ \right\}$
     and $\left\{ a_-, a_-^+ \right\}$ are two commuting
sets of $qp$-bosons.   More precisely,  from (1) we have
$$
\eqalign{
& a_+a^+_+ \; - \; p a^+_+ a_+ \; = \; q^{N_1} \qquad
  a_+a^+_+ \; - \; q a^+_+ a_+ \; = \; p^{N_1}        \cr
& a_-a^+_- \; - \; p a^+_- a_- \; = \; q^{N_2} \qquad
  a_-a^+_- \; - \; q a^+_- a_- \; = \; p^{N_2}        \cr
& [a_+, a_-]     \; = \;
  [a^+_+, a^+_-] \; = \;
  [a_+, a^+_-]   \; = \;
  [a^+_+, a_-]   \; = \; 0
}
\eqno (3)
$$
where the number operators $N_1$ and $N_2$ are defined via
$$
N_i |n_1 \rangle \otimes
    |n_2 \rangle \; = \; n_i
    |n_1 \rangle \otimes
    |n_2 \rangle \quad (i = 1,2)
\eqno (4)
$$
as in the nondeformed case.

By introducing
$$
n_1 = j + m \qquad n_2 = j - m \qquad n_1 \in \grn \qquad n_2 \in \grn
\eqno (5{\rm a})
$$
and
$$
|jm           \rangle \, \equiv \,
|j + m, j - m \rangle \,
= \, |n_1     \rangle \otimes
     |n_2     \rangle \; \in \;
{\cal F}_1 \otimes {\cal F}_2
\eqno (5{\rm b})
$$
equation (1) can be rewritten in the form
$$
\eqalign{
a_+   \; |jm \rangle \; = \;
& {\sqrt{[[j + m + {1 \over 2} - {1 \over 2}]]_{qp}}}\;
|j - {1\over 2}, m - {1\over 2} \rangle \cr
a_+^+ \; |jm \rangle \; = \;
& {\sqrt{[[j + m + {1 \over 2} + {1 \over 2}]]_{qp}}}\;
|j + {1\over 2}, m + {1\over 2} \rangle \cr
a_-   \; |jm \rangle \; = \;
& {\sqrt{[[j - m + {1 \over 2} - {1 \over 2}]]_{qp}}}\;
|j - {1\over 2}, m + {1\over 2} \rangle \cr
a_-^+ \; |jm \rangle \; = \;
& {\sqrt{[[j - m + {1 \over 2} + {1 \over 2}]]_{qp}}}\;
|j + {1\over 2}, m - {1\over 2} \rangle
\cr
}
\eqno (6)
$$
{}From equation (6),  we  see  that  we can
construct bilinear forms of the operators
$a_+$, $a_+^+$, $a_-$, and $a_-^+$
which behave like step operators on the $j$'s and/or $m$'s.
Indeed, the ($qp$-deformed spherical angular momentum) operators
$$
J_- \; = \; a^+_- a_+ \quad
J_3 \; = \; {1 \over 2} \left( N_1 - N_2 \right) \quad
J_0 \; = \; {1 \over 2} \left( N_1 + N_2 \right) \quad
J_+ \; = \; a^+_+ a_-
\eqno (7)
$$
satisfy
$$
\eqalign{
  J_- \; |jm \rangle \; = \; &{\sqrt {[[j + m]]_{qp} \;
    [[j - m + 1]]_{qp}}} \; |j, m-1 \rangle\cr
  J_3 \; |jm \rangle \; = \; &m \; |jm \rangle \quad \quad \qquad \quad
  J_0 \; |jm \rangle \; = \;  j \; |jm \rangle \cr
  J_+ \; |jm \rangle \; = \; &{\sqrt {[[j - m]]_{qp} \;
    [[j + m + 1]]_{qp}}} \; |j, m+1 \rangle\cr
}
\eqno (8)
$$
Hence, the commutators of the operators
$J_-$,
$J_3$,
$J_0$, and
$J_+$ are
$$
[J_0,J_\alpha] \; = \; 0                               \qquad
[J_3,J_\pm   ] \; = \; \pm J_\pm                       \qquad
[J_+,J_-     ] \; = \; (qp)^{J_0-J_3} \; [[2J_3]]_{qp}
\eqno (9)
$$
where $\alpha = -,3,+$.

In a similar way, the
($qp$-deformed hyperbolic angular momentum)
operators
$$
K_- \; = \; a_+a_- \qquad
K_3 \; = \; {1 \over 2} \; (N_1 + N_2 + 1)
    \; \equiv \; J_0 + {1 \over 2} \qquad
K_+ \; = \; a^+_+ a^+_-
\eqno (10)
$$
satisfy
$$
\eqalign{
 K_- \; |jm \rangle \; = \; &{\sqrt {[[j-m + {1 \over 2} - {1 \over 2}]]_{qp}
                                  \, [[j+m + {1 \over 2} - {1 \over 2}]]_{qp}}}
 \; |j-1, m \rangle \cr
 K_3 \; |jm \rangle \; = \; &(j+{1 \over 2})
 \; |j    m \rangle \cr
 K_+ \; |jm \rangle \; = \; &{\sqrt {[[j-m + {1 \over 2} + {1 \over 2}]]_{qp}
                                  \, [[j+m + {1 \over 2} + {1 \over 2}]]_{qp}}}
 \; |j+1, m \rangle \cr
}
\eqno (11)
$$
The commutators of the
operators $K_-$, $K_3$, $J_3$, and $K_+$ are
$$
\eqalign{
[J_3, K_\alpha]   = 0   \qquad
[K_3, K_{\pm} ] & = \pm
      K_{\pm}           \cr
[K_+, K_-]        = - [[2K_3]]_{qp}&+ (1-qp)[[K_3 + J_3 - {1 \over 2}]]_{qp}
                                            [[K_3 - J_3 - {1 \over 2}]]_{qp}
}
\eqno (12)
$$
from which we recognize  the Lie brackets
of u$_{1,1}$ when $q = p^{-1} \to 1$.

Equations (9) and (12) are the starting point for generating the quantum
algebras (as Hopf algebras) $U_{qp}({\rm u}_    2)$ and
                            $U_{qp}({\rm u}_{1,1})$,
respectively. Note that we can form other bilinears, in the
$qp$-boson operators, in addition to (7) and (10)~; this leads
to the quantum algebra $U_{qp}({\rm o}_{3,2})$ which is of
special relevance for studying the ``Wigner-Racah'' algebras of
 $U_{qp}({\rm u}_    2)$ and
 $U_{qp}({\rm u}_{1,1})$ [2,3].
We shall focus here on the algebra
$U_{qp}({\rm u}_2)$. For the applications, it is enough to
mention that the co-product $\Delta_{qp}$
of $U_{qp}({\rm u}_2)$ is defined by
$$
\eqalign{
 \Delta_{qp}(J_\pm) \; = \;
&J_\pm \otimes (qp)^{{1\over 2}J_0} (qp^{-1})^{+{1\over 2}J_3} +
               (qp)^{{1\over 2}J_0} (qp^{-1})^{-{1\over 2}J_3}
 \otimes J_\pm\cr
 \Delta_{qp}(J_3) \; = \;
&J_3 \otimes I + I \otimes J_3 \qquad \quad
 \Delta_{qp}(J_0) \;= \; J_0 \otimes I + I \otimes J_0 \cr
}
\eqno (13)
$$
and that the operator
$$
C_2 (U_{qp} ({\rm u}_2)) \; = \;
  {1 \over 2} \; (J_+ J_- + J_- J_+) +
  {1 \over 2} \; [[2]]_{qp} \; (qp)^{J_0-J_3} \; [[J_3]]^2_{qp}
\eqno (14)
$$
is an invariant of $      U_{qp} ({\rm u}_2)  $.
The eigenvalue  of the Casimir
                   $ C_2 (U_{qp} ({\rm u}_2)) $ on the subspace
$\left\{ \vert jm \rangle : m = -j, -j+1, \cdots, j \right\}$
is simply $[[j]]_{qp} [[j+1]]_{qp}$. The quantum algebra
$U_{qp}({\rm u}_2)$ is a  two-parameter  quantum algebra.
Note that the hermitean conjugation property
$J_-^\dagger = J_+$ requires
that either $q$ and $p$ are real or $p = \bar q$. The algebra
$U_{qp}({\rm u}_2)$
clearly leads to the ``classical'' quantum algebra
$U_{q}({\rm su}_2)$ when $p = q^{-1}$ and
$U_{q}({\rm su}_2)$ yields in turn the Lie algebra su$_2$ when $q \to 1$.

We may wonder whether we
really gain something when passing from
the ``classical'' quantum algebra
$U_q({\rm su}_2)$ to the quantum algebra $U_{qp}({\rm u}_2)$.
In this connection, let us define the operators $A_\alpha$
($\alpha = -,3,0,+$) through
$$
J_\pm \, = \, (qp)^{{1\over 2}(A_0 - {1\over 2})} \, A_\pm
  \qquad \quad
  J_0 \, = \, A_0
  \qquad \quad
  J_3 \, = \, A_3
\eqno (15)
$$
and let us introduce
$$Q = (qp^{-1})^{1\over 2}
  \qquad \quad
  P = (qp)^{1\over 2}
\eqno (16)
$$
Then, we can verify that the set
$\{A_-, A_3, A_+\}$ spans $U_Q({\rm su}_2)$, which commutes
  with $A_0$, so that we have central extension
$$
U_{qp}({\rm  u}_2) =
       {\rm  u}_1 \otimes
   U_Q({\rm su}_2)
\eqno (17)
$$
On the other hand, the invariant $C_2(U_{qp}({\rm u}_2))$
can be  developped as
$$
  C_2(U_{qp}({\rm u}_2)) \; = \;
  P^{2A_0-1} \; C_2(U_Q({\rm su}_2))
\eqno (18)
$$
where
$$
C_2(U_Q({\rm su}_2)) \; = \;
  {1 \over 2} \; (A_+A_- + A_-A_+) +
  {1 \over 2} \; [2]_Q \; [A_3]^2_Q
\eqno (19)
$$
is an invariant of $U_Q({\rm su}_2)$. Therefore,
in spite of the fact that the transformation (15-16)
allows us to generate the one-parameter algebra $U_Q({\rm su}_2)$ from the
two-parameter algebra $U_{qp}({\rm u}_2)$, the invariant
$C_2(U_{qp}({\rm u}_2))$
given by (18) still exhibits two independent parameters
($Q$ and $P$). For physical applications, the introduction of
a second parameter gives more flexibility in fitting procedures
and/or phenomenological approaches.

To close this section, it should be mentioned that
multi-parameter (in parti\-cular
  two-parameter)
quantum algebras have been studied by many authors
(see for example Refs.~[4-7]).

\sa
\sb

\centerline {\bf 3. Application to dynamical systems}

\sb

An important preliminary step for applying $q$-quantization
processes is to know $q$- and/or $qp$-analogues of ordinary dynamical
systems. We shall be interested here in nonrelativistic dynamical
systems corresponding to a charged particle
(of reduced mass $\mu=1$)
embedded in a scalar potential $V$. The case of a 4-potential,
involving a vector potential (corresponding to an Aharonov-Bohm
situation, or a monopole or a dyonium), can be addressed in a
similar way.

Among the various dynamical systems used in physics,
the oscillator system in $\gr$
and
the Coulomb system in $\gr^3$
are two paradigms of considerable importance. We shall briefly
discuss $qp$-analogues for the latter two systems and for three
parent systems
[viz., the Smorodinsky-Winternitz (SW) system,
the generalized oscillator             system and
the generalized Coulomb                system].
In the following, we use units such that $\hbar = 1$.

1. The oscillator system. The oscillator system
in $\gr^N$ is a superposition of one-dimensional oscillator
systems corresponding to potentials of type
$V = {1 \over 2} \Omega^2 x^2$
with $\Omega > 0$. Such a system is maximally superintegrable
with $2N-1$ constants of motion. For $N=1$, the
$qp$-quantization of the oscillator system may be achieved
by extending (to $p \ne q^{-1}$)
the recipe given independently by many authors
(see Refs.~[8-13]). The energy spectrum for the
one-dimensional $qp$-deformed oscillator so-obtained reads
$$
E = {1 \over 2} \> \Omega \> \left( [[n]]_{qp} + [[n+1]]_{qp}
                             \right) \qquad n \in \grn
\eqno (20)
$$
Note that $E$ is real if $q$ and $p$ are real or if
$p = \bar q$. By using equation (16), formula (20) can be rewritten
as
$$
E = {1 \over 2} \> \Omega \> P^n \left( {1 \over P} [n]_{Q} + [n+1]_{Q}
                                 \right)
\eqno (21)
$$
Two particular cases are of special interest when
$Q=q$ and $P=1$~: For $p^{-1} = q
= {\rm e}^{\psi}$            (with $\psi \in \gr$), we have
$$
E = {1 \over 2} \> \Omega \> { { \sinh (2n+1) {\psi \over 2} } \over
                               { \sinh        {\psi \over 2} } }
\eqno (22)
$$
while for $p^{-1} = q
= {\rm e}^{{\rm i} \varphi}$ (with $\varphi \in \gr$), we obtain
$$
E = {1 \over 2} \> \Omega \> { { \sin  (2n+1) {\varphi \over 2} } \over
                               { \sin         {\varphi \over 2} } }
\eqno (23)
$$
The energy $E$ as given by (22) or (23) can be expanded in
terms of the nondeformed eigenvalue $(\Omega/2) (n + 1/2)$.

2. The Coulomb system. The attractif Coulomb
system in $\gr^3$ corresponds to the potential
$ V = \alpha (1/r) $ with
$\alpha < 0$. This system is maximally superintegrable
with five constants of motion. By applying the
Kustaanheimo-Stiefel transformation (i.e., the Hopf fibration
$S^3 \to S^2$ of compact fiber $S$), we can transform the
$\gr^3$ Coulomb system into a coupled
pair of $\gr^2$ oscillator systems. The
$qp$-quantization of the Coulomb system may thus be accomplished
by $qp$-quantizing the $\gr^2$ oscillator systems [14].
We thus obtain a $qp$-analogue of the Coulomb system in $\gr^3$
for which the discrete energy spectrum is
$$
E   =   {1 \over \nu^2} E_0  \qquad
E_0 = - {1 \over 2} \alpha^2 \qquad
\nu = {1 \over 4} \sum_{i=1}^4 \left(
         [[n_i]]_{qp} + [[n_i + 1]]_{qp}
                               \right)
\eqno (24)
$$
It should be noticed that a
similar $qp$-quantization process can be effectuated for the
Coulomb system in $\gr^5$ by using the Hopf fibration
$S^7 \to S^4$ of compact fiber $S^3$.

3. The Smorodinsky-Winternitz system. The SW system
in $\gr^N$ may be considered as a superposition of one-dimensional
systems corresponding to potentials of type
$$
V = {1 \over 2} \Omega^2 x^2 + {1 \over 2} P {1 \over x^2}
\eqno (25)
$$
where $\Omega > 0$ and $P > 0$. The SW system was originally
introduced for $N=2$ [15]. For $N=3$, the SW
potential is of the $V_1$ type in the classification of
Ref.~[16]~; this potential allows the separation
of variables in the Schr\"odinger equation
in eight systems of coordinates [17].
For $N$ arbitrary, the SW system is maximally superintegrable
with $2N-1$ constants of motion [17].
Going back to $N=1$, we may $qp$-quantize
the SW system by using the approach developed
in [18,19]. The energy spectrum for the
$qp$-deformed SW system so-obtained is discrete only and given by
$$
E = \Omega \left(  [[n]]_{qp} + [[n+1]]_{qp} + \sqrt{ {1 \over 4} + P }
           \right) \qquad n \in \grn
\eqno (26)
$$
In the case where $p^{-1} = q \to 1$, the energy
$E$ reduces to the one for the nondeformed one-dimensional SW
system [15,19]
(note the sign in front of the square root, cf. Refs.~[17,20]).
It should be observed that,
in the limiting situation for which $p^{-1} = q = 1$ and $P=0$,
corresponding to the ordinary oscillator system, we must
replace (26) by $E = \Omega (n + n + 1 \pm {1 \over 2}) =
                     \Omega (k           + {1 \over 2})$
where $k$ may be equal to $2n+1$ or $2n$.

4. The generalized oscillator system. This system corresponds in
$\gr^3$ to the potential
[in circular cylindrical coordinates ($\rho, \varphi, z$)]
$$
V = {1 \over 2} \Omega^2 (\rho^2 + z^2) +
    {1 \over 2} P {1 \over z^2   }      +
    {1 \over 2} Q {1 \over \rho^2}
\eqno (27)
$$
where $\Omega > 0$, $P > 0$, and $Q > 0$. The
potential (27) is of the $V_3$ type in the classification
of Ref.~[16]. It allows the separation of
variables in the Schr\"odinger equation
in four systems of coordinates (spherical,
circular cylindrical,  problate spheroidal, and
                         oblate spheroidal coordinates).
The three-dimensional generalized
oscillator system is minimally superintegrable
with four constants of motion.
By using the approach of Ref.~[19], we can
derive a $qp$-analogue for this system. Its
energy spectrum is given by
$$
\eqalign{
   E = 2 \Omega \nu                           \qquad
&\nu = {1 \over 2} \sum_{i=1}^2   \left(
[[n_i  ]]_{qp} +
[[n_i+1]]_{qp} + \vert S_i \vert \right)      \cr
  \vert S_1 \vert&= \sqrt { m^2         + Q } \qquad
  \vert S_2 \vert = \sqrt { {1 \over 4} + P } \cr
& n_1 \in \grn                                \qquad
  n_2 \in \grn                                \qquad
  m   \in \grz
}
\eqno (28)
$$
and is discrete only.

5. The generalized Coulomb system. This system corresponds in
$\gr^3$ to the potential
[in spherical coordinates ($r, \theta, \varphi$)]
$$
V = \alpha {1 \over r} +
    \beta  { {\cos \theta} \over {r^2 \sin^2 \theta} } +
    \gamma { 1             \over {r^2 \sin^2 \theta} }
\eqno (29)
$$
where $\alpha < 0$ and $ \gamma \ge \vert \beta \vert $. The
potential (29) is of the $V_4$ type in the classification of
Ref.~[16]. It allows the separation of
variables in the Schr\"odinger equation
in spherical and parabolic coordinates.
The three-dimensional generalized
Coulomb system is minimally superintegrable
with four constants of motion.
By using the approach of Ref.~[18], we can derive a
$qp$-analogue for this system. Its discrete energy spectrum
is
$$
\eqalign{
  E   = & {1 \over \nu^2} \> E_0                  \qquad
  E_0 = - {1 \over 2    } \> \alpha^2             \qquad
  \nu =   {1 \over 2    } \> \sum_{i=1}^2 \left(
[[n_i  ]]_{qp} +
[[n_i+1]]_{qp} + \vert S_i \vert       \right)           \cr
        & \vert S_i \vert
      = \sqrt { m^2 + 2 [\gamma + (-1)^i \beta] } \qquad
n_1 \in \grn                                      \qquad
n_2 \in \grn                                      \qquad
m   \in \grz
}
\eqno (30)
$$
Note that the occurrence of a similar quantum number $\nu$ in
(28) and (30) is re\-mi\-niscent of the well-known connection
between harmonic oscillator system and Coulomb system.

For each of the $qp$-deformed sytems 1 to 5, in the limiting
situation for which $p^{-1} = q = 1$, we recover the spectra
corresponding to the nondeformed systems. The case where $q$
and $p$ are close to 1 may be used for mimicking some
perturbation effects. In this respect, let us consider the
example of the hydrogen atom and of its $q$-analogue (we take
$p^{-1} = q$). We know that the level for the principal quantum
number $n=2$ (i.e., $\ell=0$ and $\ell=1$) exhibits a fourfold
(or eightfold, if spin is taken into account) degeneracy
corresponding to the subspace $2s (\ell=0) \oplus
                               2p (\ell=1)$. In the
$q$-quantization picture, it can be shown from (24) that
the $n=2$ level splits into two doublets (or quartets,
if spin is taken into account). Furthermore, the obtained
level splitting exactly reproduces the Dirac splitting of the
$n=2$ level, namely,
$(2p \; ^2P_{3 \over 2}) -
 (2s \; ^2S_{1 \over 2} \> , \>
  2p \; ^2P_{1 \over 2})$, when
$$
q = 1 + {1 \over { \sqrt{3} } } \alpha
\eqno (31)
$$
where $\alpha$ stands here for the fine structure
constant. We have here an application of type 1-2.

\sa
\sb

\centerline {\bf 4. Application to chemical elements}

\sb

Let us go now to an application of type 2-3.
Atoms and ions can be builded from the filling,
with some prescription (taking into account the Pauli
exclusion principle), of the various $n \ell$ shells of
the hydrogen atom. Neutral atoms are reasonably
well-described by the ordering
$$
\eqalign{
1s < 2s < 2p &< 3s < 3p < 4s < 3d < 4p < 5s                         \cr
             &< 4d < 5p < 6s < 4f < 5d < 6p < 7s < 5f < 6d < \cdots
}
\eqno (32)
$$
while positive monatomic ions correspond to the sequence
$$
\eqalign{
1s < 2s < 2p &< 3s < 3p < 3d < 4s < 4p < 4d                        \cr
             &< 5s < 5p < 4f < 5d < 6s < 6p < 5f < 6d < 7s < \cdots
}
\eqno (33)
$$
The filling of the atomic shells is thus different for atoms
and ions. For instance, for the neutral atom Ti(I) we have the
atomic configuration
$
1s^2 \> 2s^2 \> 2p^6 \> 3s^2 \> 3p^6 \> 4s^2 \> 3d^2
$
and for the tripositive ion Ti(IV) the filling is
$
1s^2 \> 2s^2 \> 2p^6 \> 3s^2 \> 3p^6 \> 3d^1
$.

We want to describe here an Aufbau Prinzip based on~:  (i) the
use of the ${\rm so}_4$ symmetry of the hydrogen atom, (ii) the breaking of the
           ${\rm so}_4$ symmetry via an
           ${\rm so}_3$ invariant term, (iii) the replacement of the chain
           ${\rm so}_4 \supset
            {\rm so}_3$ by the chain
           ${\rm so}_4 >
        U_q({\rm so}_3)$, and (iv) the filling of the $n \ell$ shells arising
from       ${\rm so}_4 >
        U_q({\rm so}_3)$ according to the Pauli
principle.

Let us first briefly describe how the chain
${\rm so}_4 \supset
 {\rm so}_3$
occurs in this problem. By using the Fock stereographic projection, we
know how to express the Hamiltonian $H$ of the hydrogen atom as a
function of         the Hamiltonian $\Lambda^2$
[the eigenvalues of which are $\lambda(\lambda+2)$
with $\lambda \in \grn$]
for the four-dimensional symmetric rotor. In convenient units,
the operator $H$ reads
$$
H = - {1 \over 2} { 1 \over {\Lambda^2 + 1} }
\eqno (34)
$$
whose eigenvalues are $ - (1/2) (1/n^2) $, where
$n=\lambda+1$ is the principal quantum number. Following Novaro
[21], we may think to break the ${\rm so}_4$ symmetry by replacing
$\Lambda^2$ by $\Lambda^2 + \alpha L^2$, where the asymmetry
parameter $\alpha$ is real and $L^2$ is the
Casimir operator of ${\rm so}_3$
[the eigenvalues of which are $\ell(\ell+1)$ with $\ell \in \grn$].
The replacement of the symmetric rotor by an asymmetric one
thus introduces the orbital quantum number $\ell$. Then, the
energy of the $n \ell$ shell is
$$
E = - {1 \over 2} { 1 \over {n^2 + \alpha \ell (\ell + 1)} }
\eqno (35)
$$
It is known that $ \alpha = 4/3 $ reproduces in a
reasonable way the ordering (32) for neutral atoms [21]. However,
there exists no value of $\alpha$ for reproducing in an acceptable
way the ordering (33) for positive ions.

The next step is to $q$-quantize the chain
${\rm so}_4 \supset
 {\rm so}_3$.
The minimal extension of the Novaro model is obtained by
substituting the quantum algebra
$U_q({\rm so}_3)$ to the Lie algebra
    ${\rm so}_3 $ [22]. This leads to a new Hamiltonian whose eigenvalues are
given by (35) with the substitutions
$\ell (\ell+1) \mapsto [\ell]_q[\ell+1]_q$ and
       $\alpha \mapsto \alpha(q)$. The dependence in $q$ is
introduced not only at the level of the Casimir operator of
$U_q({\rm so}_3)$ but also in the parameter $\alpha$. A simple model
is obtained for $ \alpha = 3 - (5/3) q $ which gives back
                $ \alpha =      4/3    $ for $q=1$. The
ordering of the $n \ell$ shells is thus controlled by the
expression
$$
n^2 + (3 - { 5 \over 3 } q ) \> [\ell]_q \> [\ell+1]_q
\eqno (36)
$$
{}From the latter expression, we obtain a good classification
of~: (i) neutral atoms for $q = 0.9$, (ii) positive monoatomic
ions for $1.1 < q < 1.4$, and (iii) hydrogenlike ions for
$1.4 < q < 1.8$. Note that the hydrogen atom corresponds to the
limiting value $q=9/5$. It is to be emphasized that the
so-obtained
classification of atoms is better than the one afforded by
the Novaro model (that corresponds to $q=1$).

\sa
\sb

\centerline {\bf 5. Application to rotational spectroscopy}

\sb

As a third application
(indeed, an application of type 2),
we now describe a model for rotational
spectroscopy of molecules and nuclei. This model is based upon
the Hamiltonian
$$
H \;=\; { 1 \over 2{\cal I} } \; C_2 (U_{qp}({\rm u}_2)) + E_0
\eqno (37)
$$
where $E_0$ is some constant (e.g.,  the bandhead energy for a
deformed or superdeformed nucleus) and ${\cal I}$ denotes the
moment of inertia of the nucleus or molecule under
study. The diagonalization of $H$ within a subspace of
constant angular momentum $J$ (a spin angular momentum for a
nucleus or a molecular angular momentum for a molecule)
leads to the energies
$$
  E \; = \; { 1 \over {2 {\cal I}} } \;
  [[J]]_{qp} \; [[J + 1 ]]_{qp} + E_0
\eqno (38)
$$
or equivalently
$$
  E \; = \; { 1 \over {2{\cal I}} } \;
      {\rm e}^{ (2J-1) {{s+r} \over 2} } \;
      { {\sinh      (J {{s-r} \over 2}) \;
         \sinh [(J+1)  {{s-r} \over 2}]} \over
        {\sinh^2 (     {{s-r} \over 2})} }
      + E_0
\eqno (39)
$$
where we have introduced $s = \ln q$ and
      $r = \ln p$. For evident reasons, $E$ should be real.
Therefore, we can take either
$ (s-r) \in         \gr $ and
$ (s+r) \in         \gr $
or
$ (s-r) \in {\rm i} \gr $ and
$ (s+r) \in         \gr $. In the case
$ (s-r) \in {\rm i} \gr $ and
$ (s+r) \in         \gr $, by introducing
$$
  \matrix{
  {{s+r}\over 2        } = \beta \cos \gamma\cr
\cr
  {{s-r}\over 2 {\rm i}} = \beta \sin \gamma\cr
  }
  \quad
  \matrix{
  \Longleftrightarrow \cr
  }
  \quad
  \matrix{
  q = {\rm e}^{\beta \cos \gamma} \; {\rm e}^{+{\rm i} \beta \sin \gamma}\cr
\cr
  p = {\rm e}^{\beta \cos \gamma} \; {\rm e}^{-{\rm i} \beta \sin \gamma}\cr
  }
\eqno (40)
$$
(where $\beta$ and $\gamma$ are two independent real
parameters), the spectrum of $H$ is given by
$$
E \; = \; { 1 \over {2 {\cal I}} } \;
      {\rm e}^{(2J-1) \beta \cos \gamma} \;
      { {\sin (J \beta \sin \gamma) \; \sin [(J+1) \beta \sin \gamma]} \over
        {\sin^2 (\beta \sin \gamma)} }
      + E_0
\eqno (41)
$$
Similarly, in the case
$ (s-r) \in \gr $ and
$ (s+r) \in \gr $, by putting
$$
  \matrix{
  {{s+r}\over 2        } = \beta \cos \gamma\cr
\cr
  {{s-r}\over 2        } = \beta \sin \gamma\cr
  }
  \quad
  \matrix{
  \Longleftrightarrow \cr
  }
  \quad
  \matrix{
  q = {\rm e}^{\beta \cos \gamma} \; {\rm e}^{+ \beta \sin \gamma}\cr
\cr
  p = {\rm e}^{\beta \cos \gamma} \; {\rm e}^{- \beta \sin \gamma}\cr
  }
\eqno (42)
$$
(where here again $\beta$ and $\gamma$ are real),
the eigenvalues of $H$ are
$$
E \; = \; { 1 \over {2 {\cal I}} } \;
      {\rm e}^{(2J-1) \beta \cos \gamma} \;
      { {\sinh     (J \beta \sin \gamma) \;
         \sinh [(J+1) \beta \sin \gamma]} \over
        {\sinh^2 (    \beta \sin \gamma)} }
      + E_0
\eqno (43)
$$
Both equations (41) and (43) can be rewritten in the form
$$
E = { 1 \over {2{\cal I}_{\beta \gamma}} }    \left(
\sum_{n=0}^\infty
      d_n (\beta,\gamma) [C_2 ({\rm su}_2)]^n
                     + [2 C_1 ({\rm  u}_1) + 1]
\sum_{n=0}^\infty
      c_n (\beta,\gamma) [C_2 ({\rm su}_2)]^n \right) + E_0
\eqno (44)
$$
where
$$
{\cal I}_{\beta \gamma} =
  {\cal I} \, {\rm e}^{2 \beta \cos \gamma} \quad \qquad
  C_2 ({\rm su}_2) =
  J(J+1) \quad \qquad
  C_1 ({\rm u}_1) = J
\eqno (45)
$$
The expansion coefficients $c_n(\beta, \gamma)$
                             and $d_n(\beta, \gamma)$ in (44)
are given by series involving special functions.

The model inherent to formula (38) gives back the rigid
rotor model in the limiting situation where $p = q^{-1} = 1$.
The model
corresponding to $p^{-1} = q = {\rm e}^{ {\rm i} \beta }$
($\beta \in \gr$) was introduced by Raychev {\it et al.} [23]
for describing rotational bands of deformed and superdeformed
nuclei. The more general two-parameter model corresponding to
$q = \bar p = {\rm e}^{         \beta \cos \gamma}
              {\rm e}^{ {\rm i} \beta \sin \gamma}$
[see formula (41)] has been successfully applied to some
superdeformed bands of even-even nuclei in the
$A \sim 190$ region [24]~; it has been shown in Ref.~[24] that
the introduction of a second parameter of a ``quantum algebra''
nature increases the agreement between theory and experiment in a
significant way. Some tests for the application
of the two-parameter model
(39) [in the versions (41) and (43)] to
molecules are presently under consideration.

\sa
\sb

\centerline {\bf 6. Concluding remarks}

\sb

{}From the applications described in sections
2 to 5, we can make the following comments.

They are several ways to
obtain a $q$- (or $qp$)-quantization
of a given dynamical system.

\item{(i)} We may start from the connection (if it is
known) between this system and oscillator systems, for
which there is a (generally well accepted) consensus
on the way to $q$-quantize them.

\item{(ii)} Another approach consists in
replacing the dynamical invariance (Lie)
algebra $g$ of the considered system by
a quantum algebra $U_q(g)$.

\item{(iii)} We can also try to convert the Schr\"odinger
(or Dirac) equation for the dynamical system into an equation
involving $q$-derivative.

Of course, there is no reason to obtain the same $q$-quantized
system from the approaches (i) to (iii). It is even possible to
obtain two different $q$-quantized systems when working inside
a given approach. (This is the case for the hydrogen atom for
example.)

Similar remarks may be done about the derivation of a
$q$-analogue of a given physical model.

The lack of unicity in deriving $q$-deformed objects is
obviously a source of pessimism in applications of quantum
algebras to physics.

Another major drawback is the impossibility to obtain a
universal significance of the deformation parameter $q$. For
instance, $q$ may be connected to the fine structure constant
for the hydrogen atom [14], to the softness parameter in
rotational spectroscopy of nuclei [23], and to the chemical
potential in Bose-Einstein statistics [25]. Furthermore,
the parameter $q$, although useful from a phenomenological
point of view, may have no physical significance. This is the
case for the classification of chemical elements [22] or for
the formation of coherent structures in strongly interacting
$q$-boson systems [26].

Finally, in many cases, the
results afforded by a $q$-quantization of a given model can
be equally well obtained from an extension (out of the quantum
algebra context) of the model.

The balance between optimism and pessimism seems to go towards
pessimism. ``What is the use of quantum groups~?~[27]'' That is
the question we have to face.

\sa
\sb

\centerline {\bf References}

\sb
\baselineskip = 0.55 true cm

\noindent
\item{[1]} Kibler, M.,
in {\it Symmetry and Structural
Properties of Condensed Matter}, ed. W. Florek, D. Lipinski
and T. Lulek, World Scientific~: Singapore (1993).

\noindent
\item{[2]} Kibler, M. and Grenet, G.,
J. Math. Phys. {\bf 21} (1980) 422.

\noindent
\item{[3]} Smirnov, Yu.F. and Kibler, M.R.,
in {\it Symmetries in Science VI~: From the Rotation Group to
Quantum Algebras}, ed. B. Gruber, Plenum~: New York (1993).

\noindent
\item{[4]} Sudbery, A.,
J. Phys. A {\bf 23} (1990) L697.
% multiparametre: [alg\`ebre quantique \`a deux param\`etres]

\noindent
\item{[5]}
Chakrabarti, R. and Jagannathan, R.,
J.~Phys.~A {\bf 24} (1991) L711.

\noindent
\item{[6]} Schirrmacher, A., Wess, J. and Zumino, B.,
Z.~Phys.~C {\bf 49} (1991) 317.
% pq deformation : two-parameter deformation of GL(2)

\noindent
\item{[7]} Fairlie, D.B. and Zachos, C.K.,
Phys.~Lett.~{\bf 256B} (1991) 43.
% [multi-parameter deformations]

\noindent
\item{[8]} Arik, M. and Coon, D.D.,
J.~Math.~Phys.~{\bf 17} (1976) 524.
% deformed bosons (a a+ -  q a+ a = 1)
% parle d'etats coherents pour l'oscillateur harmonique
% q-deforme (a la facon des matheux: a a* - q a* a = 1)

\noindent
\item{[9]} Kuryshkin, V.,
Ann.~Fond.~Louis de Broglie {\bf 5} (1980) 111.
% deformed bosons (a a+ - mu a+ a = 1)

\noindent
\item{[10]} Jannussis, A., Brodimas, G., Sourlas, D., Vlachos,
K., Siafarikas, P. and Papaloucas, L., Hadronic J.~{\bf 6} (1983) 1653.
% deformed bosons : A A+ - q A+ A = 1 and hadronic oscillator

\noindent
\item{[11]} Macfarlane, A.J.,
J.~Phys.~A {\bf 22} (1989) 4581.
% deformed bosons and boson realization of su_q(2)

\noindent
\item{[12]} Biedenharn, L.C.,
J.~Phys.~A {\bf 22} (1989) L873.
% deformed bosons and boson realization of su_q(2)

\noindent
\item{[13]}
Katriel, J. and Kibler, M.,
J.~Phys.~A {\bf 25} (1992) 2683.

\noindent
\item{[14]} Kibler, M. and N\'egadi, T.,
J.~Phys.~A {\bf 24} (1991) 5283.
% atomic spectroscopy

\noindent
\item{[15]}
Fri\v s, J., Mandrosov, V., Smorodinsky, Ya.A., Uhl\'\i \v r, M.
and Winternitz, P.,
Phys.~Lett.~{\bf 16} (1965) 354.
% SW system

\noindent
\item{[16]}
Makarov, A.A., Smorodinsky, J.A., Valiev, Kh. and Winternitz, P.,
Nuovo Cimento A {\bf 52} (1967) 1061.

\noindent
\item{[17]} Evans, N.W.,
J. Math. Phys. {\bf 31} (1990) 600.
% Evans = ``Group theory of the Winternitz system''

\noindent
\item{[18]} Dr\u ag\u anescu, Gh.E., Campigotto, C. and Kibler, M.,
Phys. Lett. {\bf 170A} (1992) 339.
% SYTEME ABC

\noindent
\item{[19]} Kibler, M. and Campigotto, C.,
Phys.~Lett.~{\bf 181A} (1993) 1.
% SYSTEME ABO

\noindent
\item{[20]} Bonatsos, D., Daskaloyannis, C. and Kokkotas, K.,
{\it Deformed oscillator algebras for two-dimensional
     quantum superintegrable systems}, to be published.

\noindent
\item{[21]} Novaro, O.,
Int.~J.~Quantum Chem.~{\bf S7} (1973) 53.

\noindent
\item{[22]} N\'egadi, T. and Kibler, M.,
J.~Phys.~A {\bf 25} (1992) L157.
% atomic spectroscopy

\noindent
\item{[23]}
\noindent Raychev, P.P., Roussev, R.P. and Smirnov, Yu.F.,
J. Phys. G {\bf 16} (1990) L137.
% [$SU_q(2)$ and rotational spectra of deformed nuclei]

\noindent
\item{[24]}
\noindent Barbier, R., Meyer, J. and Kibler, M.,
J. Phys. G~: in press.
% [rotational spectra of deformed nuclei]

\noindent
\item{[25]} Tuszy\'nski, J.A., Rubin, J.L., Meyer, J. and Kibler, M.,
Phys. Lett. {\bf 175A} (1993) 173.
% STATISTIQUE

\noindent
\item{[26]} Tuszy\'nski, J.A. and Kibler, M.,
J.~Phys.~A {\bf 25} (1992) 2425.
% coherent structures in condensed matter physics

\noindent
\item{[27]} Fairlie, D.B.,
{\it What is the use of quantum groups}, to be published.

\bye